\documentclass[a4paper,11pt]{article}

\newcommand{\vecx}[1]{\hspace{-0.1mm}\vec{\hspace{0.1mm}#1}}

\usepackage{jheppub}     
\usepackage[T1]{fontenc} 

\usepackage{enumerate}
\usepackage{graphicx}
\usepackage{caption}
\usepackage{subcaption}
\usepackage{amsmath}
\usepackage{braket}
\usepackage{amsmath}
\usepackage{slashed}
\usepackage{mathrsfs}
\usepackage{simplewick}
\usepackage{tensor}
\usepackage{bbm}
\usepackage{multicol}
\usepackage[compat=1.1.0]{tikz-feynman}
\usepackage{placeins}
\usepackage{wrapfig}
\usepackage{array}
\usepackage{multirow}
\usepackage{amssymb}
\usepackage{epstopdf}
\usepackage{graphicx}
\usepackage{contour}

\title{\boldmath Complete $\mathcal{O}(\alpha_s^2)$ Corrections to the
                 Leptonic Invariant Mass Spectrum in $b\to X_c l\bar{\nu}_l$ Decay}

\author[a]{Mateusz Czaja,}
\author[a]{Miko{\l}aj Misiak,}
\author[b,c]{Abdur Rehman}

\affiliation[a]{Institute of Theoretical Physics, Faculty of Physics, University of Warsaw,\\
                02-093 Warsaw, Poland.}
\affiliation[b]{Department of Physics, University of Alberta, Edmonton, AB T6G 2J1, Canada.}
\affiliation[c]{Department of Environmental and Physical Sciences, Faculty of Science,\\
                    Concordia University of Edmonton, Alberta, T5B 4E4, Canada.}

\emailAdd{mp.czaja@uw.edu.pl}
\emailAdd{misiak@fuw.edu.pl}
\emailAdd{rehman3@ualberta.ca}

\abstract{ In the determination of the Cabibbo-Kobayashi-Maskawa
matrix element $|V_{cb}|$ from inclusive semileptonic $B$-meson
decays, moments of the leptonic invariant mass spectrum constitute
valuable observables. To evaluate them with sufficient precision,
perturbative $\mathcal{O}(\alpha_s^2)$ corrections to the analogous
spectrum in the partonic $b\to X_c l\bar{\nu}_l$ decay are necessary.
In the present paper, we compute such perturbative corrections in a
complete manner, including contributions from the
triple-charm channel, namely from the $cc\bar{c}l\bar{\nu}_l$ final
states. We present our results in terms of numerical fits
in both the single- and triple-charm
cases. We confirm the recently found results for the
single-charm correction, and analyze the triple-charm channel
impact on centralized moments of the spectrum.}

\begin{document} 
\begin{flushright}
ALBERTA-THY-10-24
\end{flushright}
\maketitle
\flushbottom

\section{Introduction} \label{sec:intro}

The Cabibbo-Kobayashi-Maskawa (CKM) matrix element $|V_{cb}|$ enters
the Standard Model (SM) predictions for many phenomenologically
important observables. The well-known examples are the
$B_s\to\mu^+\mu^-$ decay branching ratio or the neutral Kaon
mixing parameter $\epsilon_K$. The uncertainty of $|V_{cb}|$
determination in the SM limits the power of such observables in
constraining parameter spaces of beyond-SM
theories. Currently, around 50\% of the theoretical uncertainty
in both $\mathcal{B}(B_s\to\mu^+\mu^-)$~\cite{Czaja:2024the} and
$|\epsilon_K|$~\cite{Jwa:2023uuz} is due to $|V_{cb}|$, making it a
crucial element at the high-luminosity frontier of particle
physics.

The value of $|V_{cb}|$ is usually extracted from a fit to
kinematic moments of spectra in the inclusive semileptonic
$B\to X_cl\bar{\nu}_l$ decay, with $l=e,\mu$. Traditionally, either the charged
lepton energy ($E_l$) or the hadronic invariant mass squared ($r^2$)
spectra were used~\cite{Alberti:2014yda}, as these were the only
distributions initially measured at the $B$-factories
(see, e.g., ref.~\cite{BaBar:2014omp}). The first measurements of the leptonic invariant
mass squared ($q^2$) spectrum\footnote{
with low $E_l$ removed via cuts on $q^2$ only}
by Belle~\cite{Belle:2021idw} and Belle-II~\cite{Belle-II:2022evt}
opened a new chapter in the efforts to improve the precision of the SM
prediction for $|V_{cb}|$.

A comprehensive introduction to the treatment of inclusive
semileptonic $B$-meson decays in the Heavy Quark
Expansion (HQE) formalism can be found in chapter~6 of
ref.~\cite{Manohar:2000dt}. Both the decay rate and the
above-mentioned kinematic moments of spectra can be evaluated in terms
of a double power series in $\alpha_s$ and $\bar{\Lambda}/m_b$, where
$\bar{\Lambda}$ is the dominant contribution to the difference between
the $B$-meson and $b$-quark masses, i.e. $\bar{\Lambda} \simeq m_B-m_b$.
In the case of the decay rate (zeroth moment) $\Gamma_{sl} \equiv 
\Gamma(B\to X_cl\bar{\nu}_l)$, one finds\footnote{
$G_F$ denotes the Fermi constant measured in the muon decay~\cite{MuLan:2010shf}.
Throughout the paper, we neglect non-leading corrections in $\alpha_{em}$ or $m_b/$(electroweak scale).}
\begin{equation}
\label{OPE}
\Gamma_{sl}=\Gamma_{sl}^{\rm part} +
\frac{G_F^2 m_b^5|V_{cb}|^2}{192\pi^3}\sum\limits_{k}\frac{C_k\braket{O_k^{(n_k)}}}{m_b^{n_k-3}},
\end{equation}
where the partonic rate $\Gamma_{sl}^{\rm part} \equiv \Gamma(b\to
X_c^{\rm part}l\bar{\nu}_l)$ is the perturbatively calculable
inclusive decay rate of a free $b$-quark into the lepton-antineutrino
pair and any partonic state $X_c^{\rm part}$. The dominant
contribution to the r.h.s.\ of eq.~(\ref{OPE}) is given by
$\Gamma_{sl}^{\rm part}$. The remainder is a correction of order
$(\bar{\Lambda}/m_b)^2$ that is parameterized by diagonal matrix
elements (expectation values) of dimension-$n_k$
%
%
local operators $O^{(n_k)}_k$ ($n_k \geq 5$) in the decaying $B$-meson
state at rest. Explicit expressions for those operators up to
dimension 6 can be found, e.g., in eqs.~(8)-(11) of
ref.~\cite{Mannel:2021zzr}. The matrix elements $\braket{O_k^{(n_k)}}$
scale like $\bar{\Lambda}^{n_k-3}$, i.e.,
\begin{align}
\braket{O_k^{(n_k)}} \equiv
\frac{\braket{B(\vecx{p}=0)|O^{(n_k)}_k\!(x)|B(\vecx{p}=0)}}{2m_B}
~\sim~
\bar{\Lambda}^{n_k-3},
\end{align}
under the standard normalization convention
$\;\braket{B(\vecx{p})|B(\vecx{q})} = 2 m_B (2\pi)^3
\delta^{(3)}(\vecx{p}-\vecx{q})$. Their $x$-independence is easily
verified by writing $O^{(n_k)}_k\!(x) = e^{iPx}\, O^{(n_k)}_k\!(0)\,
e^{-iPx}$, and then using the fact that $\ket{B(\vecx{p})}$ is an
eigenstate of the momentum operator $P$.


The structure of the local operators $O_k^{(n_k)}$ is
obtained in the operator product expansion but values of their
matrix elements are non-perturbative quantities that have to be
extracted together with $|V_{cb}|$ in a fit to experimental
data. The dimensionless Wilson coefficients $C_k$
parameterize short-distance QCD effects. They are
functions of $\frac{m_c}{m_b}$ only, and can be perturbatively
determined, order-by-order in $\alpha_s$.

Expressions analogous to eq.~(\ref{OPE}) hold for kinematic
moments of spectra, too. Therefore, one can constrain the dominant
matrix elements $\braket{O_k^{(n_k)}}$ by using experimental information
on such moments.  Moreover, as shown in ref.~\cite{Mannel:2018mqv},
certain sets of non-perturbative matrix elements come in fixed
linear combinations in Reparametrization-Invariant (RPI)
observables. Consequently, less operators need to be treated as
independent. Up to $n_k=6$, there are 4 generic operators,
and 3 operators needed for RPI observables.\footnote{
Provided one neglects the $n_k=6$ operators whose Wilson
coefficients vanish at the tree level --- see section~5
ref.~\cite{Mannel:2018mqv}.}
Up to $n_k=7$, these numbers are 13 and 8, respectively.  A
spectral moment is RPI provided the kinematic variables
of the considered spectrum are independent of the $B$-meson
four-velocity $v$. Moments of the $q^2$ spectrum are
therefore RPI, while those in $E_l=vp_l$ or
$r^2=(m_Bv-p_l-p_{\bar\nu_l})^2$ are not.


The experimental setups usually veto events with small $E_l$ due to
large backgrounds. On the theory side, removing $E_l<E_{cut}$
would break the RP invariance, necessitating the inclusion of all
possible operators. Instead, one can remove the low $E_l$ region with
a lower cut on $q^2$ -- see figure~6.2 of
  ref.~\cite{Manohar:2000dt}. The $q^2$ spectrum with a cut
  remains RPI, and preserves its simpler non-perturbative
structure, simultaneously being compatible with the
experimental requirements. The first fits using the $q^2$
spectrum were already performed in
refs.~\cite{Bernlochner:2022ucr,Finauri:2023kte}.

To calculate moments of the $q^2$ spectrum with a cut, one needs to
begin with the dominant, perturbative terms in the formulae that are
analogous to eq.~(\ref{OPE}). At any given order in $\alpha_s$, it is
sufficient to determine $d\Gamma_{sl}^{\rm part}/dq^2$, and then
evaluate its moments with any experimentally desirable cut. Such a
calculation has recently been completed up to
$\mathcal{O}(\alpha_s^2)$ in ref.~\cite{Fael:2024gyw} without
including numerically subdominant contributions from the triple-charm
channel, namely from the $cc\bar{c}l\bar{\nu}_l$ final states.

In the present work, we complete the calculation of the
$\mathcal{O}(\alpha_s^2)$ correction to $d\Gamma_{sl}^{\rm part}/dq^2$
by including the triple-charm contributions. We also independently
confirm the single-charm results of ref.~\cite{Fael:2024gyw}.
Our results are provided in terms of precise numerical fits for
the spectrum as a function of $q^2$, the charm quark mass $m_c$,
and the renormalization scale $\mu$ at which the strong coupling
constant is renormalized.

Let us note that corrections to $d\Gamma_{sl}/dq^2$ that are
suppressed by $(\bar{\Lambda}/m_b)^2$ and $(\bar{\Lambda}/m_b)^3$
have already been determined up to $\mathcal{O}(\alpha_s^1)$
in refs.~\cite{Mannel:2021zzr,Alberti:2012dn,Alberti:2013kxa}.
Moreover, both $\Gamma_{sl}^{\rm part}$ and the first few
moments of $d\Gamma_{sl}^{\rm part}/dq^2$ without any cut are
already known up to $\mathcal{O}(\alpha_s^3)$ from
refs.~\cite{Fael:2020tow} and~\cite{Fael:2022frj}, 
respectively. However, no moment measurements without cuts are
available.

The article is organized as follows. In section~\ref{sec:method}, we
discuss details of our calculation. In section~\ref{sec:res},
we present results for the $\mathcal{O}(\alpha_s^2)$ 
correction to $d\Gamma_{sl}^{\rm part}/dq^2$ and its
moments, with subsections \ref{sec:3c} and \ref{sec:1c} dedicated,
respectively, to the triple-charm $b\to X_{3c}\, l\bar\nu_l$ 
and single-charm $b\to X_{1c}\, l\bar\nu_l$ 
channels, and subsection \ref{sec:moms} describing the impact 
on $q^2$ moments. We conclude in section~\ref{sec:conclusions}. 
Various numerical fit coefficients are collected in
appendix~\ref{sec:app}. In the ancillary file, we provide these fits
in a \texttt{Mathematica}-readable format.

\section{Details of the calculation} \label{sec:method}

\subsection{The $q^2$-spectrum}

In the standard first step of the calculation, all particles
heavier than the $b$-quark are integrated out of the SM. As
already mentioned, we work at the leading order in
$\alpha_{em}$ and $m_b/$(electroweak scale). In such a case,
one is left with a single effective operator, namely a
Fermi-like interaction between the quark and lepton currents. The
leptonic current contribution factorizes, and is fixed to
all orders in $\alpha_s$. Consequently, the unpolarized
differential rate of the semileptonic $b$-quark decay can be
written as
\begin{equation}
\label{diffRate}
d\Gamma^{\rm part}_{sl}=\frac{1}{2}\frac{1}{2m_b}dPS_{p_b\to p_l p_{\bar\nu_l} p_c \{ p_X \} }\frac{1}{M_W^4}\left[\sum_X\sum_{\sigma_H}|\mathcal{M}_{b\to W^*cX}|^2\right]_{\mu\nu}\left[\sum_{\sigma_l}|\mathcal{M}_{W^*\to l\bar\nu_l}|^2\right]^{\mu\nu} .
\end{equation}
Here, $X$ stands for all the final-state partons except for the
charm quark that gets produced in the $b$-quark annihilation
vertex. As mentioned above, the matrix element factorizes
into the hadronic and leptonic parts, denoted by $\mathcal{M}_{b\to
W^*cX}$ and $\mathcal{M}_{W^*\to l\bar\nu_l}$ respectively, summed
over the discrete degrees of freedom indexed with $\sigma_H$ and
$\sigma_l$. The phase space element can be factorized as~\cite{Aquila:2005hq}
\begin{equation}
\label{PSfact}
dPS_{p_b\to p_l p_{\bar\nu_l} p_c \{ p_X \} }=\frac{1}{128\pi^4m_b} dE_l dq^2 dr^2 dPS_{r\to p_c \{ p_X \} },
\end{equation} 
where $q\equiv p_l+p_{\bar\nu_l}$ and $r\equiv p_b-q$. Throughout this
work, we treat $l$ and $\bar\nu_l$ as massless.

In the notation of eq.~(\ref{diffRate}), the leptonic and partonic tensors are given by
\begin{equation}
\label{tensorDef}
\begin{split}
g_{ew}^2 L_{\mu\nu} &\equiv \left[\sum_{\sigma_l}|\mathcal{M}_{W^*\to l\bar\nu_l}|^2\right]_{\mu\nu},\\
g_{ew}^2|V_{cb}|^2 W_{\mu\nu} &\equiv \int dPS_{r\to p_c \{ p_X \} }\left[\sum_X\sum_{\sigma_H}|\mathcal{M}_{b\to W^*cX}|^2\right]_{\mu\nu},
\end{split}
\end{equation}
where $g_{ew}$ is the electroweak $SU(2)$ gauge coupling. Combining eqs. (\ref{diffRate})--(\ref{tensorDef}), we obtain
\begin{equation}
\label{WLspectr}
\frac{d\Gamma^{\rm part}_{sl}}{dq^2dr^2}=\frac{G_F^2|V_{cb}|^2}{16\pi^4m_b^2}W_{\mu\nu}\int\limits_{E_-}^{E_+}dE_lL^{\mu\nu},
\end{equation} 
where $E_{\pm}\equiv (q^0\pm|\vecx{q}|)/2$ are the kinematical bounds on the lepton energy $E_l$
in the decaying $b$-quark rest frame, for fixed $q^2$ and $r^2$. 

The leptonic tensor can be evaluated in a straightforward manner. It reads
\begin{equation}
L^{\mu\nu}\equiv p_l^\mu p_{\bar\nu_l}^\nu -(p_l p_{\bar\nu_l})g^{\mu\nu}+p_{\bar\nu_l}^\mu p_l^\nu -i\epsilon^{\mu\nu\rho\sigma}(p_l)_\rho (p_{\bar\nu_l})_\sigma.
\end{equation}
The partonic tensor can be decomposed into Lorentz structures as
\begin{equation}
\label{hadronic tensor decomposition}
W^{\mu\nu}=-W_1\left(q^2,r^2\right)g^{\mu\nu}+W_2\left(q^2,r^2\right) \frac{p_b^{\mu} p_b^{\nu}}{m_b^2} + \mbox{(other)},
\end{equation}
where the omitted possible terms vanish when contracted with the $E_l$
integral of $L_{\mu\nu}$ in the limit of massless leptons. Combining
the above two equations allows for a straightforward integration on
the r.h.s.\ of eq.~(\ref{WLspectr}), yielding
\begin{equation}
\label{spectrum with leptons}
\frac{d\Gamma^{\rm part}_{sl}}{dq^2}=\frac{G_F^2|V_{cb}|^2}{16\pi^4m_b^2}
\int\!dr^2\,|\vecx{q}|\left(W_1q^2+\frac{W_2}{3}|\vecx{q}|^2\right).
\end{equation}

\subsection{The effective width formula}

\begin{figure}[t]
	\centering
	\includegraphics[width=\textwidth]{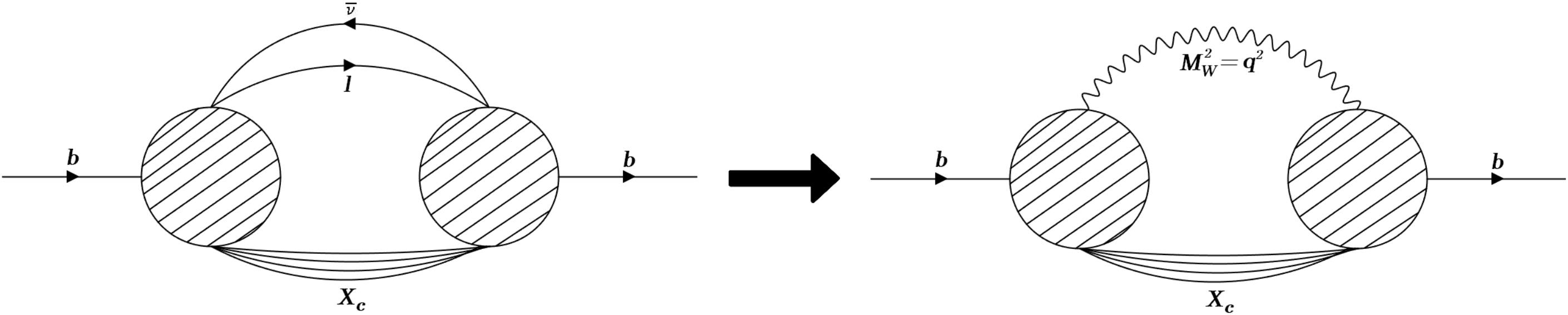}
	\caption{Diagrammatic depiction of the replacement of the lepton pair by an auxiliary vector boson.}\label{fig:eff}
\end{figure}

The r.h.s.\ of eq.~(\ref{spectrum with leptons}) is reminiscent
of the top-quark decay rate or, more generally, decay rate
$\widetilde{\Gamma}$ of a massive fermion into a lighter fermion
and a massive vector boson with four-momentum $q$, mediated
by a left-handed fermion current. To derive a relation between
$\widetilde{\Gamma}$ and $d\Gamma^{\rm part}_{sl}/dq^2$,
we can consider the SM with parameters tuned to retain the
physical values of quark masses, CKM-matrix elements and $G_F$, but
with the $W$-boson mass set to $\sqrt{q^2}$. In other words, we
can consider the SM with so small electroweak gauge couplings
that the $W$ boson is lighter than the $b$ quark, and 
$\widetilde{\Gamma} = \Gamma(b \to X_cW)$. 
To facilitate the use of the optical theorem, we choose the
unitary gauge for the light $W$-bosons. Such a choice removes all
contributions of would-be Goldstone bosons from the amplitude. As
before, we decouple all particles heavier than the $b$ quark, and
neglect non-leading electroweak corrections.

Let $p_b$, $p_c$, and $\{p_X\}$ be defined as in the previous
subsection, while $q$ now stands for the final-state $W$-boson
momentum in $b \to X_cW$. The differential phase-space element of
this process can be decomposed as
\begin{equation}
dPS_{p_b\to p_cqp_X}=\frac{dr^2}{2\pi}dPS_{p_b\to qr}dPS_{r\to p_c\{p_X\}}.
\end{equation}
The two body volume $dPS_{p_b\to qr}$ for fixed $q^2$ and $r^2$ is constant and reads
\begin{equation}
dPS_{p_b\to qr}=\frac{1}{16\pi^2}\frac{|\vecx{q}|}{m_b}d\Phi_q,
\end{equation}
where $\Phi_q$ is the angular direction of the $W$-boson. In the rest
frame of the $b$-quark, the length $|\vecx{q}|$ is fixed for given
values of $q^2$, $r^2$, and $m_b^2$. In this frame, the decay is also
isotropic, so in the total width we can simply replace
\begin{equation}
dPS_{p_b\to qr}\to\frac{1}{4\pi}\frac{|\vecx{q}|}{m_b}.
\end{equation}
The unpolarized total rate of this process can now be written as
\begin{equation}
\label{GW with W}
\widetilde{\Gamma}=\frac{q^2 G_F \sqrt{2}}{m_b}\frac{|V_{cb}|^2
}{8\pi^2 m_b} \int\! dr^2\, |\vecx{q}|\, W^{\mu\nu}
\sum_{\rho}\varepsilon^{(\rho)*}_\mu \varepsilon^{(\rho)}_\nu,
\end{equation}
where $\rho$ indexes the final-state $W$-boson polarization, and
$\varepsilon^{(\rho)}_\mu$ are the polarization vectors that satisfy
\begin{equation}
\label{pol sum rule}
\sum_{\rho}\varepsilon^{(\rho)*}_\mu\varepsilon^{(\rho)}_\nu=-g_{\mu\nu}+\frac{q_\mu q_\nu}{q^2}.
\end{equation}
In eq.~(\ref{GW with W}), the $W^{\mu\nu}$ tensor is defined exactly
as in eq.~(\ref{tensorDef}), because the non-leptonic part of each
diagram is the same in the true-SM semileptonic decay and in
the modified-SM $b \to X_cW$ decay. Using the polarization sum
rule (\ref{pol sum rule}) and the decomposition\footnote{
The omitted part on the r.h.s.\ of eq.~(\ref{hadronic tensor decomposition})
vanishes under contraction with the r.h.s.\ of eq.~(\ref{pol sum rule}).}
(\ref{hadronic tensor decomposition}) leads to the following effective width formula
\begin{equation}
\label{effective width}
\frac{d\Gamma^{\rm part}_{sl}}{dq^2}=\frac{\sqrt{2}G_F}{12\pi^2}\widetilde{\Gamma} .
\end{equation}

The replacement of the lepton pair with an auxiliary light vector
boson, schematically shown in figure~\ref{fig:eff}, has already
been applied in previous calculations of the $q^2$ spectrum ---
see, e.g., eq.~(1)\footnote{
An extra factor of $q^2/M_W^2$ there w.r.t.\ our eq.~(\ref{effective width})
is due to different notational conventions.}
of ref.~\cite{Dowling:2008ap}. It serves two
purposes. First, it allows for a straightforward use of the optical
theorem while retaining the $q^2$ dependence of the spectrum in the
vector-boson mass. If one would proceed with the calculation with the
leptons, this dependence would be lost when the matrix element is
integrated over all final states. To prevent this, one could use the
equivalent method of reverse unitarity~\cite{Anastasiou:2002yz}, but
the effective width formula derived above proves to be easier to
implement in practice. An additional benefit is that the lepton loop
is evaluated automatically. With these two simplifications, to obtain
the $q^2$ spectrum including the $\mathcal{O}(\alpha_s^n)$ correction,
one has to calculate the imaginary parts of $(n+1)$-loop diagrams with
3 independent mass scales: $m_b$, $m_c$, and $\sqrt{q^2}$.

\subsection{Calculation of the bare effective width}

\begin{figure}[t]
	\centering
	\includegraphics[width=\textwidth]{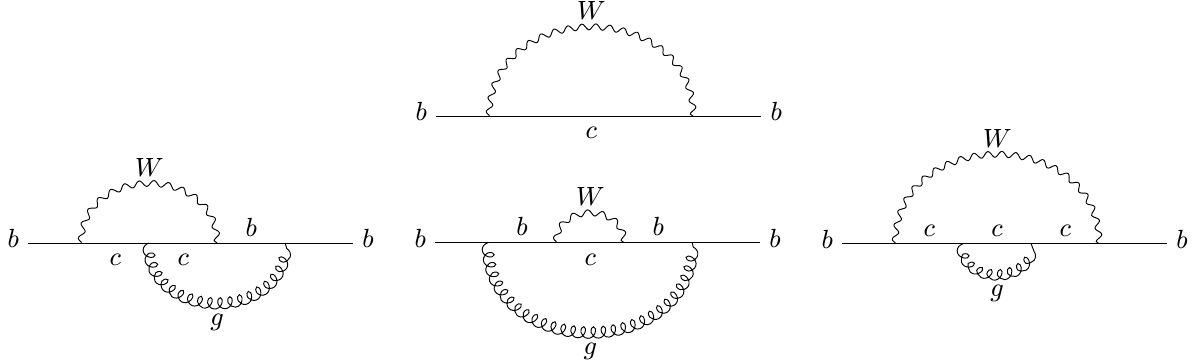}
	\caption{Feynman diagrams contributing to $\widetilde{\Gamma}$ at the LO in QCD (top) and the NLO in QCD (bottom).
	          At the NLO, the SIs were IBP-reduced to MIs belonging to the topology inherited from the first diagram
		  in the bottom row.}\label{fig:NLOdiags}
\end{figure}

We compute the auxiliary width $\widetilde{\Gamma}$ using the optical theorem
\begin{equation}
	\widetilde{\Gamma}=\frac{1}{2}\frac{1}{2m_b}2\Im\left[ -i\left(\sum\limits_{\sigma_b}
	\begin{tikzpicture}[baseline=-\the\dimexpr\fontdimen22\textfont2\relax]
		\begin{feynman}[inline=(o), large]
			\vertex [blob] (o) {};
			\vertex [left=1.5cm of o] (b1) {\(b\)};
			\vertex [right=1.5cm of o] (b2) {\(b\)};			
			\diagram* {
				(b1) -- (o), (o) -- (b2)
			};
		\end{feynman}
	\end{tikzpicture}
	\right)\right].
\end{equation}
Before turning to the calculation of $\mathcal{O}(\alpha_s^2)$ terms, we used
eq.~(\ref{effective width}) to compute the known Leading Order (LO)
$q^2$ spectrum, and the Next-to-LO (NLO) QCD correction. We generated
the necessary $b$-propagator diagrams using \texttt{QGRAF}~\cite{Nogueira:1991ex}, and
obtained the resulting algebraic expressions using a custom
\texttt{Mathematica} code. The LO and NLO diagrams are shown in
figure~\ref{fig:NLOdiags}. Next, we expressed the contribution of each
diagram to the amplitude as a linear combination of Scalar Integrals
(SIs). They were reduced to a set of Master Integrals (MIs) using
Integration-By-Parts (IBP) identities implemented in
\texttt{KIRA}~\cite{Maierhofer:2017gsa,Klappert:2020nbg}. Both at the
LO and NLO in QCD, the SIs were reduced to MIs from single
topologies. For the evaluation of the MIs at the LO and NLO, we
utilized the Differential Equations (DEs) in the canonical form method
with the help of the \texttt{Mathematica} package
\texttt{LIBRA}~\cite{Lee:2020zfb}, and expressed the resulting
Goncharov polylogarithms as dilogarithms. The boundary conditions for
the DEs were determined numerically using the auxiliary mass flow
method~\cite{Liu:2017jxz,Liu:2021wks}, implemented in the
\texttt{Mathematica} package \texttt{AMFlow}~\cite{Liu:2022chg}. Next, they were
reconstructed as combinations of integers and $\pi^2$ using the
\texttt{Mathematica} implementation of the PSLQ algorithm.

\begin{figure}[t]
	\centering
	\includegraphics[width=\textwidth]{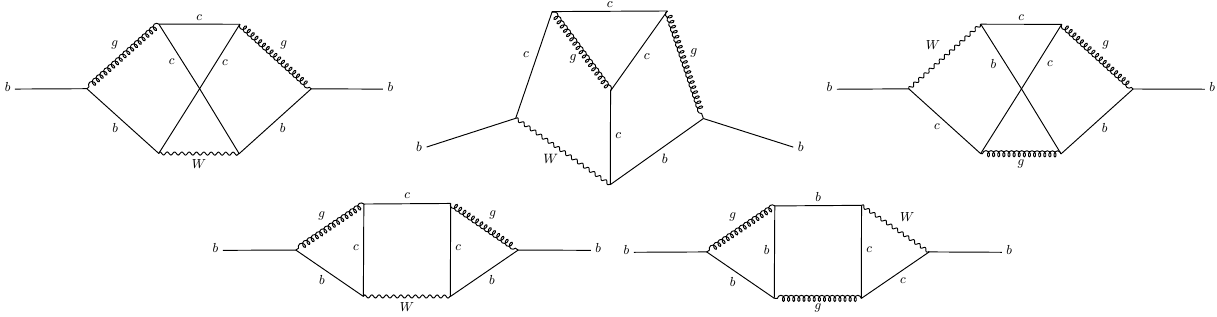}
	\caption{Sample diagrams contributing to $\widetilde{\Gamma}$ at the NNLO in QCD.
	         All SIs at this order were IBP-reduced to MIs from topologies inherited from
		 the displayed five diagrams.}\label{fig:NNLOdiags}
\end{figure}

At the Next-to-NLO (NNLO), we identified 39 bare diagrams, with some
examples shown in figure~\ref{fig:NNLOdiags}. We proceeded with the
same method as at lower orders, performing IBP reductions of all SIs
to obtain 98 MIs belonging to five different topologies.  However, contrary
to our approach at the LO and NLO, we did not use the DE
method to evaluate them. Instead, we computed all the NNLO MIs
using \texttt{AMFlow} at 682 different points in the
$(q^2/m_b^2,m_c^2/m_b^2)\equiv(\hat{q}^2,\hat{m}_c^2)$ space,
obtaining a dense scan of the bare part of $\widetilde{\Gamma}$.

\subsection{Renormalization}

The bare part of $\widetilde{\Gamma}$ has been
computed using the dimensional regularization method in
$D\equiv4-2\epsilon$ dimensions . After taking the imaginary
part of the MIs, the LO contribution is finite, the NLO correction has
only a simple $\epsilon$ pole, and at the NNLO in QCD there are at
most $1/\epsilon^2$ divergences. Not all of these divergences are
purely ultraviolet, even though the quantity we consider
is infrared- and collinear-safe. To arrive at the finite result, we
perform the standard renormalization. We renormalize the quark fields
and masses in the on-shell scheme, and use $\alpha_s$ in the
$\overline{\text{MS}}$ scheme with five active flavours.

The renormalized $\widetilde{\Gamma}$ is obtained by combining
the bare and counterterm contributions as follows:
\begin{equation}
\widetilde{\Gamma}= \lim_{\epsilon\to0}\left[\widetilde{\Gamma}^{\rm bare}+\widetilde{\Gamma}^{\otimes}\right]\equiv
\lim_{\epsilon\to0}\left[\sum_{n=0}\alpha_s^n\widetilde{\Gamma}^{{\rm bare}\,(n)}+
                         \sum_{n=1}\alpha_s^n\widetilde{\Gamma}^{\otimes\,(n)}\right].
\end{equation}
The individual contributions of one- and two-loop diagrams with counterterm insertions yield the following renormalization formulas:
\begin{equation}
\begin{split}
\widetilde{\Gamma}^{\otimes(1)} &=R_1^{(0)} \widetilde{\Gamma}^{{\rm bare}\,(0)},\\
\widetilde{\Gamma}^{\otimes(2)} &=R_2^{(0)} \widetilde{\Gamma}^{{\rm bare}\,(0)} + R_2^{(1)} \widetilde{\Gamma}^{{\rm bare}\,(1)},
\end{split}
\end{equation}
where
\begin{equation}
\begin{split}
R_1^{(0)}&\equiv\delta_b^{(1)}+2\delta_{m_c}^{(1)}\hat{m}_c^2\partial_{\hat{m}_c^2},\\
R_2^{(0)}&\equiv\delta_b^{(2)}+\left(2\delta_{m_c}^{(2)}+2\delta_b^{(1)}\delta_{m_c}^{(1)}+\left(\delta_{m_c}^{(1)}\right)^2\right)\hat{m}_c^2\partial_{\hat{m}_c^2}+2\left(\delta_{m_c}^{(1)}\right)^2\hat{m}^4_c\partial^2_{\hat{m}_c^2},\\
R_2^{(1)}&\equiv R_1^{(0)}+2\delta_g^{(1)}+2\delta_{m_b}^{(1)}m_b^2\partial_{\widetilde{m}_b^2}\bigg\rvert_{\widetilde{m}_b=m_b},
\end{split}
\end{equation}
and
\begin{equation}
\delta_a\equiv Z_a-1\equiv\alpha_s\delta_a^{(1)}+\alpha_s^2\delta_a^{(2)}+\mathcal{O}(\alpha_s^3).
\end{equation}
The necessary renormalization constants $Z_a$ with $a=g,m_c,m_b$
can be found in refs.~\cite{Gross:1973id,Melnikov:2000zc,Bekavac:2007tk}.
In the definition of $R_2^{(1)}$, the mass $\widetilde{m}_b$ should be
inserted into the internal $b$-quark propagators when constructing
seeds to perform the IBP reduction for the SIs, shifting the diagrams
away from the on-shell relation $p_b^2=m_b^2\neq\widetilde{m}_b^2$.
After taking the partial derivative w.r.t.\ $\widetilde{m}_b^2$,
the $b$-quark mass can be put back equal to $\sqrt{p_b^2}$,
prior to the IBP reduction.

\section{Results} \label{sec:res}

\subsection{The $b\to X_{3c}\, l\bar\nu_l$ spectrum} \label{sec:3c}

\begin{figure}[t]
	\centering
	\includegraphics[width=\textwidth]{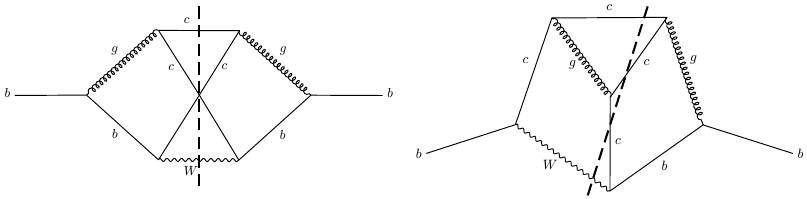}
	\caption{Two examples of cut diagrams contributing to the triple-charm channel.
	         The dashed lines indicate unitarity cuts. All the 22 cut MIs in the triple-charm calculation
		 belong to one of the two topologies inherited from the above diagrams.}\label{fig:cutdiag}
\end{figure}

Once the renormalized $\widetilde{\Gamma}$ has been computed,
we can use the 4-dimensional relation~(\ref{effective width}) to
get the semileptonic $q^2$ spectrum. The present subsection is
devoted to the triple-charm $b\to X_{3c}\, l\bar\nu_l$ partonic
channel. In the corresponding hadronic decay of the $B$-meson, the
lightest possible hadronic final states have masses of order $m_D
m_{\eta_c} \simeq 4.8\,$GeV, which differs from $m_B \simeq 5.3\,$GeV
only by ${\mathcal O}(\bar\Lambda)$. Consequently, application of the
$\bar\Lambda/m_b$ expansion as in eq.~(\ref{OPE}) becomes
questionable. We have therefore decided to separate the $b\to
X_{3c}l\bar\nu_l$ contribution from the better
controlled single-charm $b\to X_{1c}\,l\bar\nu_l$ channel,
both in the fit to the spectrum and to the moments. Such a
separation also facilitates a direct comparison with the results of
ref.~\cite{Fael:2024gyw}.

\begin{figure}[t]
	\centering
	\includegraphics[width=.8\textwidth]{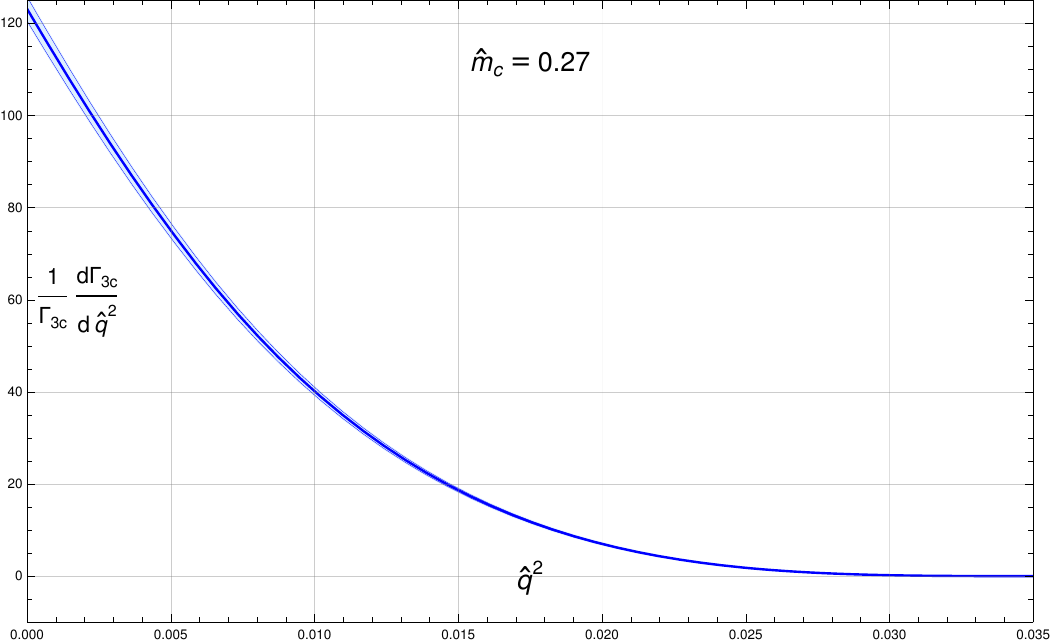}
	\caption{The $\hat{q}^2$ spectrum of the $b\to cc\bar{c}l\bar\nu_l$ decay normalized to the width of this channel.}\label{fig:3c spectr}
\end{figure}

As we are interested in the exclusive partonic $b\to
cc\bar{c}l\bar\nu_l$ decay channel, we have to consider diagrams
with unitary cuts. Out of all 3-loop MIs present in the fully
inclusive calculation, we extracted those with an allowed cut across
three $c$-quark lines. This resulted in 22 MIs belonging to 2
topologies shown in figure~\ref{fig:cutdiag}. We computed them using
\texttt{AMFlow} at 650 points in the $(\hat{q}^2,\hat{m}_c)$ space,
with the following boundaries: $\hat{m}_c=0.14$, $\hat{m}_c=0.29$,
$\hat{q}^2=0$, and $\hat{q}^2=(1-3\hat{m}_c)^2$. The last boundary
represents the largest kinematically allowed $q^2$ for the
triple-charm semileptonic decay. We obtain the numerical
results for the $q^2$ spectrum of the $b\to X_{3c}l\bar\nu_l$
channel width $\Gamma^{\rm part, 3c}_{sl}$ with 60 significant
digits. We then fit these results to the following ansatz:
\begin{equation}
\label{3c ansatz}
\frac{d\Gamma^{\rm part, 3c}_{sl}}{dq^2}=\alpha_s^2m_b^3G_F^2|V_{cb}|^2L_{3c}^{\frac{7}{2}}\left[\sum_{kl}C_{kl}\hat{q}^{2k}\hat{m}_c^l+\Theta[10L_{3c}-1]\sum_{mn}\tilde{C}_{mn}\hat{q}^{2m}\hat{m}_c^n\right]+\mathcal{O}(\alpha_s^3),
\end{equation}
where $L_{3c}\equiv\left(\hat{q}^2-(1+3\hat{m}_c)^2\right)\left(\hat{q}^2-(1-3\hat{m}_c)^2\right)$,
and the ranges of indices $k$, $l$, $m$, and $n$ are $(0,2)$, $(0,6)$, 
$(0,4)$, and $(0,6)$ respectively. The values of fit parameters
$C_{kl}$ and $\tilde{C}_{mn}$ are listed in appendix~\ref{sec:app},
tables \ref{table:3cnt} and \ref{table:3ct}, respectively. The
relative error of this ansatz when compared with the results obtained
using \texttt{AMFlow} is less than $1.5\%$ for all points in the
$(\hat{q}^2,\hat{m}_c)$ space for which we performed the numerical
calculation. The normalized spectrum of the $b\to cc\bar{c}l\bar\nu_l$
channel is shown in figure~\ref{fig:3c spectr}. The indicated
uncertainty comes entirely from the fitting error, as the renormalization scale
dependence cancels out in the normalized spectrum. This uncertainty has been
conservatively estimated by assuming a 1.5\% error over the entire fit
range. In reality, the uncertainty only reaches 1.5\% at the extreme
edge of the triple-charm phase space where the differential decay rate is
strongly suppressed for kinematical reasons.

\subsection{The $b\to X_{1c}\, l\bar\nu_l$ spectrum} \label{sec:1c}

\begin{figure}[t]
	\centering
	\includegraphics[width=.8\textwidth]{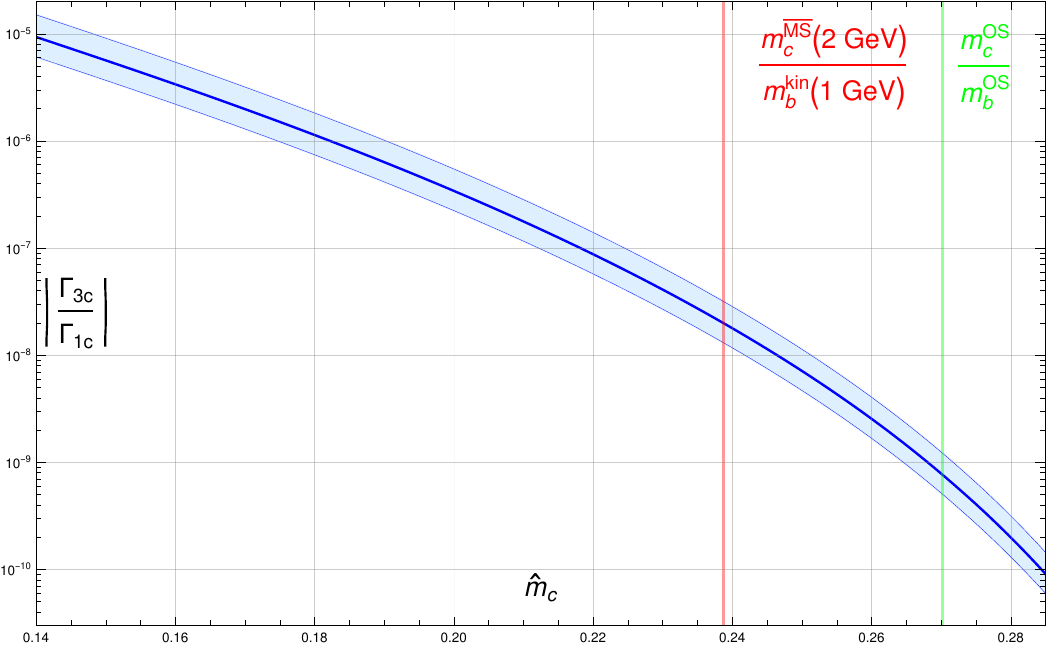}
	\caption{The ratio of the widths of the $b\to X_{3c}\,l\bar\nu_l$ and $b\to X_{1c}\,l\bar\nu_l$ channels.
	The red vertical line indicates the ratio of the $\overline{\text{MS}}$ charm mass renormalized at $2\,$GeV
	and the kinetic bottom mass with the separation scale of $1\,$GeV~\cite{Finauri:2023kte}. 
	Our reference value of the on-shell mass ratio is indicated with the green vertical line.}\label{fig:width ratio}
\end{figure}

The dominant part of the inclusive partonic width comes from the $b\to
X_{1c}\,l\bar\nu_l$ channel. The ratio of widths of the single- and
triple-charm channels as a function of $\hat{m}_c$ is shown in
figure~\ref{fig:width ratio}. The indicated theoretical
uncertainty has been estimated by varying the renormalization
scale between $2.3\,$GeV and $9.2\,$GeV. The uncertainty of the
fits in this ratio is negligible.

We compute the NNLO QCD correction to the $q^2$ spectrum of the $b\to
X_{1c}\,l\bar\nu_l$ channel by subtracting the triple-charm
contribution described in the previous subsection from the fully
inclusive correction obtained as a combination of the imaginary parts
of the 98 MIs without cuts
\begin{equation}
\frac{d\Gamma^{\rm part,1c(2)}_{sl}}{dq^2}=\frac{d\Gamma^{\rm part,(2)}_{sl}}{dq^2}
                                                 -\frac{d\Gamma^{\rm part,3c(2)}_{sl}}{dq^2}.
\end{equation}
The above subtraction is performed for numerical results obtained with
\texttt{AMFlow} with the precision of 60 significant digits at 682
points in the $(\hat{q}^2,\hat{m}_c)$ parameter space, out of which
311 points satisfied the $\hat{q}^2<(1-3\hat{m}_c)^2$ condition for a
non-vanishing triple-charm part.

Using these numerical values for the $\mathcal{O}(\alpha_s^2)$
correction to the single-charm $q^2$ spectrum, we perform a fit using
the following ansatz:
\begin{equation}
\label{1c ansatz}
\frac{d\Gamma^{\rm part,1c(2)}_{sl}}{dq^2}=m_b^3G_F^2|V_{cb}|^2L_{1c}\sum_{jkl}D_{jkl}\hat{m}_c^{j}\hat{q}^{2k}\log^l L_{1c},
\end{equation}
where
$L_{1c}\equiv\left(\hat{q}^2-(1+\hat{m}_c)^2\right)\left(\hat{q}^2-(1-\hat{m}_c)^2\right)$,
and the ranges of indices $j$, $k$, and $l$ are $(0,1)$, $(0,7)$, and
$(0,8)$ respectively. The fit coefficients $D_{jkl}$ are presented in
appendix~\ref{sec:app}, in
tables~\ref{table:1c01},~\ref{table:1c02},~\ref{table:1c11},
and~\ref{table:1c12}.

\begin{figure}[t]
	\centering
	\includegraphics[width=.8\textwidth]{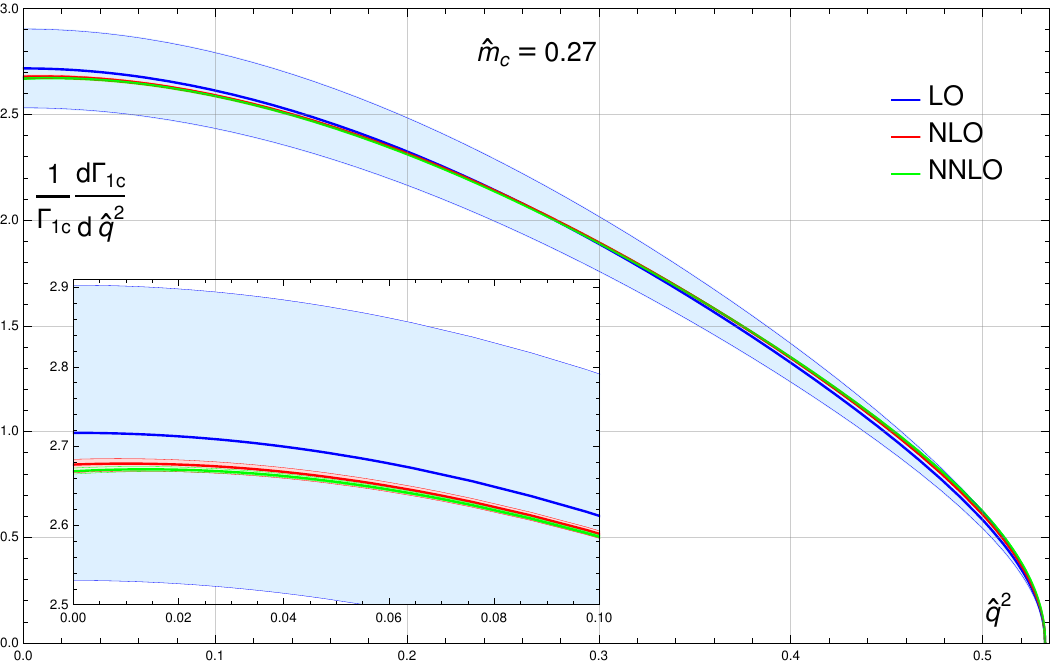}
	\caption{The $\hat{q}^2$ spectrum of the $b\to X_{1c}\,l\bar\nu_l$ decay normalized to the width of this channel.
	The bottom-left panel is the enlarged part of the spectrum close to $\hat{q}^2=0$. }\label{fig:1cSpectr}
\end{figure}

The results of the fit were compared with the analytic results
presented in ref.~\cite{Fael:2024gyw} for 13600 points, as well as
with the numerical values calculated using the method described
above. The relative error never exceeded $5.2\times10^{-4}$. The $q^2$
spectrum of the single-charm channel, including the
$\mathcal{O}(\alpha_s^2)$ correction is shown in
figure~\ref{fig:1cSpectr}. We observe a significant improvement in the
behaviour of the perturbative series for the normalized spectrum,
as compared to the unnormalized one. It suggests that a portion of
the renormalon ambiguity is mitigated when the overall factor of
$(m_b^{\text{OS}})^5$ cancels out. The uncertainty of the LO is
estimated by assuming that $\text{NLO}\approx \text{LO
}\alpha_s/\pi$. The uncertainties at the NLO and NNLO are obtained by
varying the renormalization scale in the same range as in
figure~\ref{fig:width ratio}.

\subsection{Corrections to the $q^2$-moments} \label{sec:moms}

Values of the leptonic invariant mass in the semileptonic $B$
meson decay can only be measured in bins of non-zero width. To compare
these measurements directly with theoretical predictions, one would
need to integrate the $q^2$ spectrum over the corresponding
intervals.  The HQE for such integrals is generally not
well-behaved, especially near the maximal allowed $q^2$. The
extraction of $|V_{cb}|$, the quark masses, and non-perturbative
matrix elements is instead performed by comparing theoretical
predictions and measurements for moments of the spectrum.

Focusing on the partonic approximation, let us define the
$n$-th $q^2$ moment of the semileptonic decay with a lower cut
$\hat{q}^2_{cut}$ as
\begin{equation}
\hat{M}_n(\hat{q}^2_{cut})\equiv\int\limits_{\hat{q}^2_{cut}}^{(1-\hat{m}_c)^2}d\hat{q}^2\hat{q}^{2n}\frac{d\Gamma^{\rm part}_{sl}}{d\hat{q}^2}.
\end{equation}
In the literature, $M_n=m_b^{2n}\hat{M}_n$ are often used instead.
We shall also consider normalized moments given by
\begin{equation}
\braket{\hat{q}^{2n}}_{\hat{q}^2>\hat{q}^2_{cut}}\equiv\frac{\hat{M}_n(\hat{q}^2_{cut})}{\hat{M}_0(\hat{q}^2_{cut})}.
\end{equation}
Finally, following ref.~\cite{Fael:2018vsp}, we adopt the following definition for the centralized moments:
\begin{equation}
\begin{split}
\hat{q}_1(\hat{q}^2_{cut})&\equiv\braket{\hat{q}^{2}}_{\hat{q}^2>\hat{q}^2_{cut}},\\
\hat{q}_n(\hat{q}^2_{cut})&\equiv\braket{(q^2-\braket{\hat{q}^{2}})^n}_{\hat{q}^2>\hat{q}^2_{cut}}=\\&=\sum_{j=0}^n\binom{n}{j}\braket{\hat{q}^{2j}}_{\hat{q}^2>\hat{q}^2_{cut}}(-\braket{\hat{q}^{2}}_{\hat{q}^2>\hat{q}^2_{cut}})^{n-j}.
\end{split}
\end{equation}

\begin{figure}[t]
	\centering
	\includegraphics[width=\textwidth]{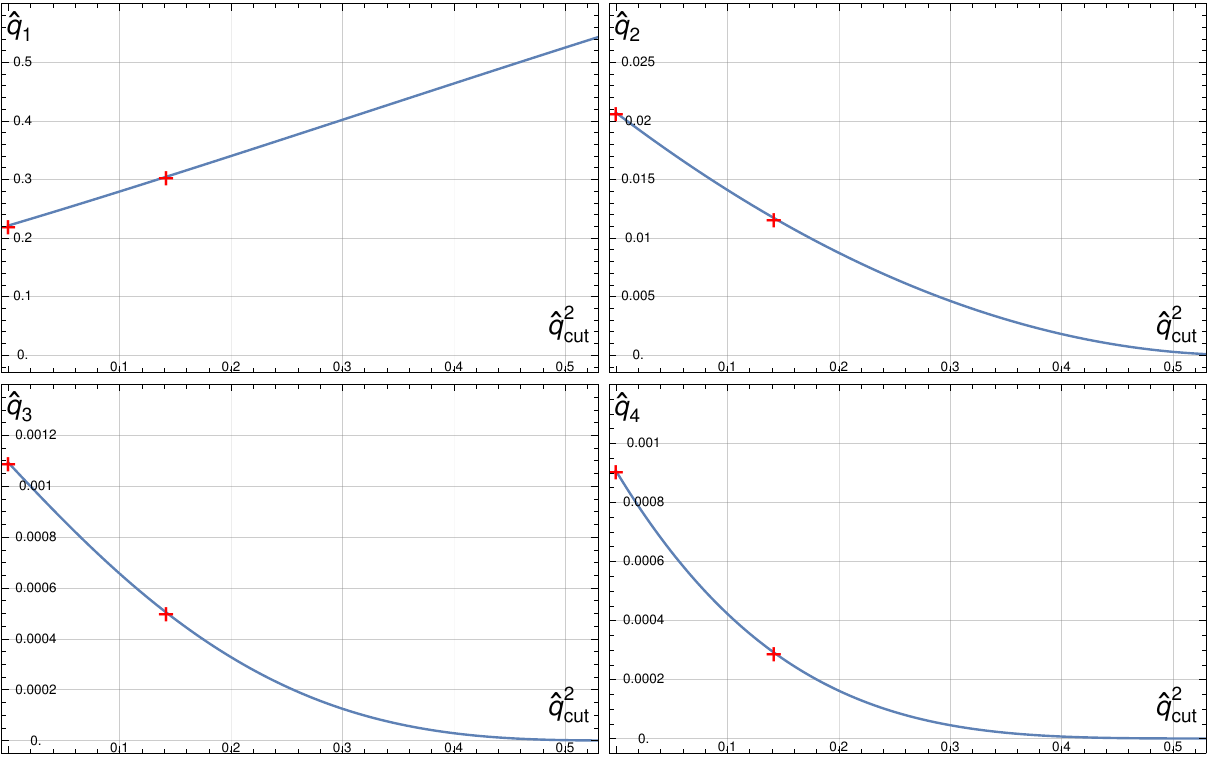}
	\caption{The first four centralized $q^2$ moments as functions of $q^2_{cut}$ for $\hat{m}_c=0.25$. With red crosses,
	we indicate the numerical results given in eqs. (21) and (22) of ref.~\cite{Fael:2024gyw}.}\label{fig:1cMoms}
\end{figure}

In figure~\ref{fig:1cMoms}, we show our results for the first four
centralized moments in the on-shell scheme as functions of
$\hat{q}^2_{cut}$ with the value of $\hat{m}_c$ set to $0.25$. To have
a direct comparison with ref.~\cite{Fael:2024gyw}, the presented centralized
moments were calculated including only the single-charm NNLO QCD
correction (\ref{1c ansatz}), and neglecting the triple-charm
channel. With red crosses, we indicate the numerical results given in
eqs. (21) and (22) of ref.~\cite{Fael:2024gyw}. We find exact
agreement to all digits given in that publication.

\begin{figure}[t]
	\centering
	\includegraphics[width=\textwidth]{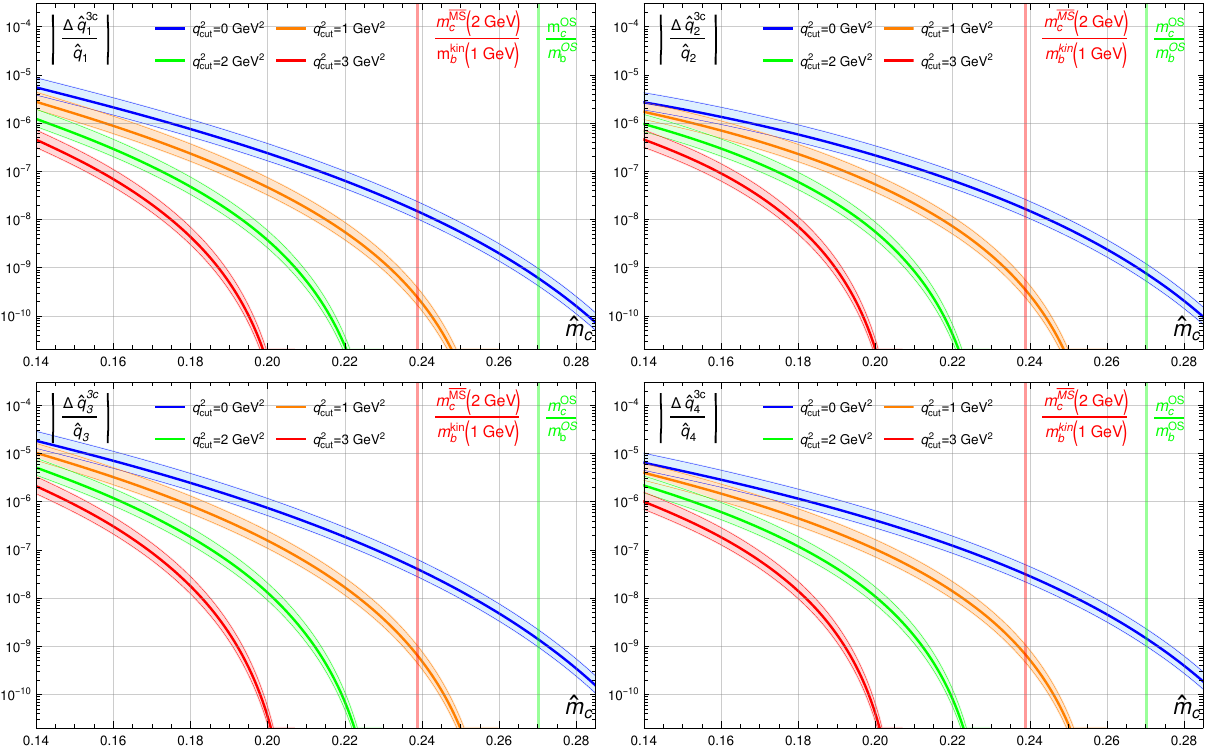}
	\caption{Relative impact of the triple-charm channel on the centralized $q^2$ moments
	         as a function of $\hat{m}_c$ and $\hat{q}^2_{cut}$. For the definition of $\Delta\hat{q}_1^{3c}$
		 see the text. The red and green vertical lines indicate the same quark mass
		 ratios as in figure~\ref{fig:width ratio}.}\label{fig:3c impact}
\end{figure}

Figure~\ref{fig:3c impact} depicts the relative impact of the
triple-charm channel on the centralized moments when compared with the
pure single-charm approximation, including the
$\mathcal{O}(\alpha_s^2)$ correction. We define $\Delta
\hat{q}^{3c}_n$ as
\begin{equation}
\Delta \hat{q}^{3c}_n\equiv\hat{q}_n-\hat{q}_n^{1c},
\end{equation}
where $\hat{q}_n^{1c}$ is the centralized moment computed up to the
NNLO QCD without including the triple-charm contribution. The above
definition can naturally be interpreted as the error of the
single-charm approximation. It decreases rapidly with increasing $\hat{m}_c$
and $\hat{q}^2_{cut}$, and for physical values of quark masses and
non-zero $q^2$ cuts it quickly becomes negligible.

\begin{figure}[t]
	\centering
	\includegraphics[width=\textwidth]{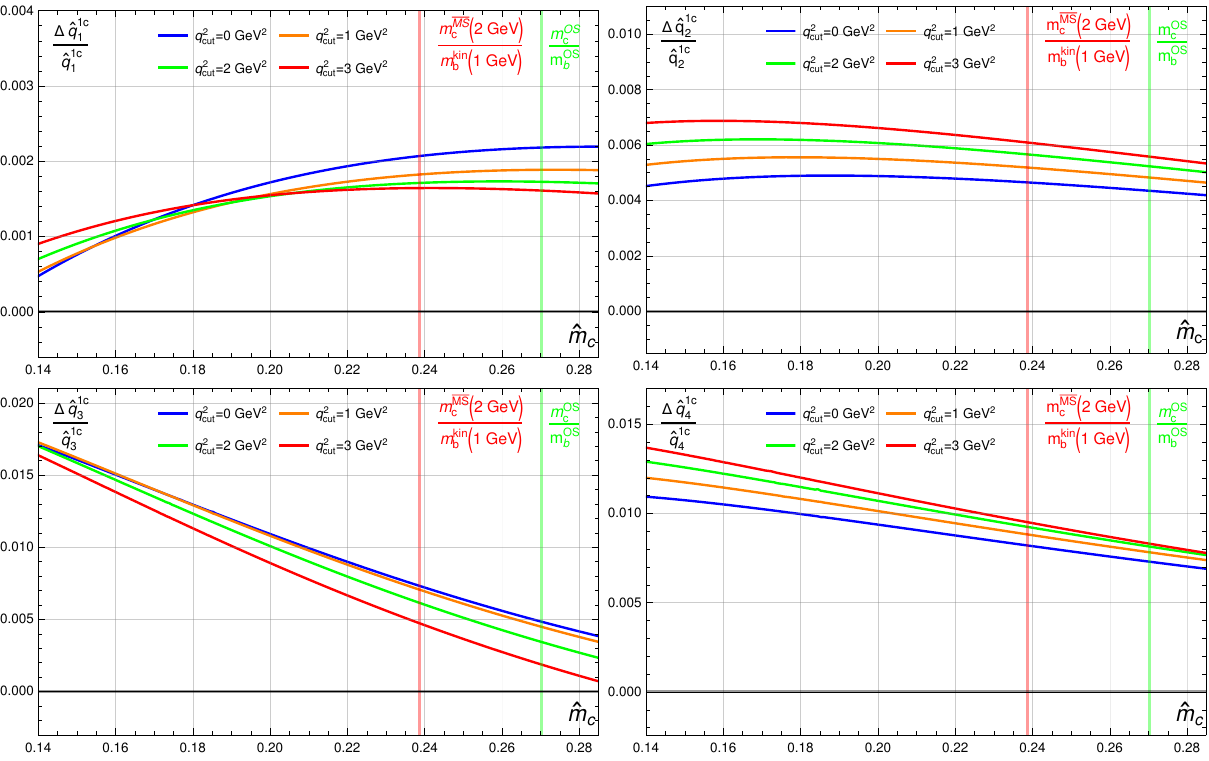}
	\caption{Relative impact of the single-charm NNLO QCD correction on the centralized $q^2$ moments
	         as a function of $\hat{m}_c$ and $\hat{q}^2_{cut}$. For the definition of $\Delta\hat{q}_1^{1c}$
		 see the text. For clarity, we omit uncertainties from varying the renormalization scale.}\label{fig:1c impact}
\end{figure}

For comparison, in figure~\ref{fig:1c impact}, we present the impact
of the single-charm NNLO QCD term on the centralized $q^2$
moments. The difference $\Delta \hat{q}^{1c}_n$ is defined as
\begin{equation}
\Delta \hat{q}^{1c}_n\equiv\hat{q}_n^{1c}-\hat{q}_n^{\text{NLO}},
\end{equation}
where $\hat{q}_n^{\text{NLO}}$ is the moment computed at the NLO only.

\section{Conclusions} \label{sec:conclusions}

The recent focus of both the theoretical and experimental communities
in the area of inclusive semileptonic decays of the $B$ meson stems
from the need for improving the precision of $|V_{cb}|$
determination in the SM. The measurements of $q^2$ moments by
Belle and Belle II, along with calculations of various perturbative
and non-perturbative effects allow to significantly enhance the
precision of extracting $|V_{cb}|$ and non-perturbative matrix
elements. This, in turn, narrows down the uncertainty of the SM
predictions for other phenomenologically important observables. The
present paper contributes to these efforts in two ways. First, it
provides an independent verification of the NNLO
single-charm contribution to the partonic $q^2$ spectrum, recently
evaluated in ref.~\cite{Fael:2024gyw}. Second, it extends this
calculation by incorporating effects of the $b\to
cc\bar{c}l\bar\nu_l$ decay channel. Our analysis made extensive
use of the auxiliary mass flow method, implemented in the
\texttt{AMFlow} package. The precise numerical results obtained with
this tool were used to perform fits to the single- and triple-charm
channel contributions to the $q^2$ spectrum, which we have
provided in the ancillary files. In addition to studying the $q^2$
spectrum itself, we analyzed the impact of the triple-charm
channel on the centralized $q^2$ moments of the spectrum, comparing it
with the dominant single-charm NNLO QCD correction. We confirm
that for the physical values of quark masses and for cuts on
low $q^2$ employed in contemporary experiments, the impact of the
triple-charm channel is minuscule. Nevertheless, our
calculation provides the first fully quantitative analysis of
this channel.

\acknowledgments

We would like to thank Matteo Fael, Paolo Gambino and Matthias
Steinhauser for useful discussions. This work was supported in part by
the National Science Center, Poland, under the research projects
2020/37/B/ST2/02746 (all the authors) and 2023/49/B/ST2/00856 (MM only).

\appendix

\section{Numerical values of fit parameters} \label{sec:app}

In this appendix, we present numerical values of the fit
coefficients for the single-charm fit, as given in eq.~(\ref{1c
ansatz}), and the triple-charm fit, as given in eq.~(\ref{3c ansatz}). The
complete expressions for the fits are available in the ancillary file.
\begin{table}[h]
\centering
\small
\begin{tabular}{c|c|c|c}
$C_{kl}$ & $k=0$ & $k=1$ &$k=2$ \\\hline
$l=0$ & -0.000862940 & -0.001237605 & -0.002930817 \\
$l=1$ & 0.022083312 & 0.058729219 & 0.115555544 \\
$l=2$ & -0.233051560 & -0.929611726 & -1.948951905 \\
$l=3$ & 1.300169847 & 7.064296877 & 17.168143423 \\
$l=4$ & -4.046951098 & -28.327414394 & -82.170972154 \\
$l=5$ & 6.666415074 & 57.939616602 & 202.387552617 \\
$l=6$ & -4.541480275 & -47.726627340 & -200.982999303 \\
\end{tabular}
\caption{Coefficients $C_{kl}$ of the triple-charm fit~(\ref{3c ansatz}).}
\label{table:3cnt}
\end{table}
\begin{table}[h]
\centering
\small
\begin{tabular}{c|c|c|c|c|c}
$\tilde{C}_{mn}$ & $m=0$ & $m=1$ & $m=2$ & $m=3$ & $m=4$\\\hline
$n=0$ & 0.000877518 & 0.001621422 & 0.003642739 & 0.004746374 & 0.640253858 \\
$n=1$ & -0.022328502 & -0.068057724 & -0.138426420 & -0.028111942 & -20.337500422 \\
$n=2$ & 0.234808565 & 1.020865993 & 2.251801608 & -1.305831462 & 258.079023134 \\
$n=3$ & -1.306653472 & -7.508277813 & -19.200975797 & 19.197488356 & -1635.101572743 \\
$n=4$ & 4.059050049 & 29.397188800 & 88.978406754 & -95.971405644 & 5172.363417944 \\
$n=5$ & -6.675376047 & -58.957345285 & -211.538554968 & 167.311712182 & -6539.450493870 \\
$n=6$ & 4.541480275 & 47.726627340 & 200.982999303 & 0 & 0 \\
\end{tabular}
\caption{Coefficients $\tilde{C}_{mn}$ of the triple-charm fit~(\ref{3c ansatz}).}
\label{table:3ct}
\end{table}
\begin{table}[h]
\centering
\small
\begin{tabular}{c|c|c|c|c}
$D_{0kl}$ & $k=0$ & $k=1$ &$k=2$ & $k=3$ \\\hline
$l=0$ &-0.00065492620361 & -0.00390985766592 & -0.00220565766436 & 0.15518605570268 \\
$l=1$ &-0.00142837487165 & 0.00696508473371 & 0.09405910531312 & -0.06253133578189 \\
$l=2$ &0.00315708546717 & -0.02325332865210 & 0.00374231943691 & -2.02104356689117 \\
$l=3$ &-0.01566245331197 & 0.00539313247413 & -0.16514145943056 & 0.50368450565789 \\
$l=4$ &-0.00314430463123 & 0.03298251534973 & -0.24771235141673 & 0.79272863991011 \\
$l=5$ &-0.00116367410900 & -0.00066823342187 & -0.01632951298570 & 0.12789891805598 \\
$l=6$ &-0.00016427141654 & -0.00086937186633 & 0.00406718033696 & 0.00295240291003 \\
$l=7$ &-0.00002838003168 & -0.00013193918044 & 0.00091118101557 & -0.00127493558773 \\
$l=8$ &-0.00000311384087 & -0.00000605460509 & 0.00008807454959 & -0.00020743224743 \\
\end{tabular}
\caption{Coefficients $D_{0kl}$ of the single-charm fit~(\ref{1c ansatz}) for $k$ between 0 and 3.}
\label{table:1c01}
\end{table}
\begin{table}[h]
\centering
\small
\begin{tabular}{c|c|c|c|c}
$D_{0kl}$ & $k=4$ & $k=5$ &$k=6$ & $k=7$ \\\hline
$l=0$ &-0.08570516624642 & -2.60307981382978 & 5.46598388371161 & -3.07155763105516 \\
$l=1$ &-4.45202557105393 & 12.27898482822888 & -12.17665279523761 & 4.05283421170514 \\
$l=2$ &6.82900780658794 & -8.98001675918627 & 4.79587351688384 & -0.71420622887211 \\
$l=3$ &-0.22696886297429 & -0.79009467604488 & 0.80044056001754 & -0.12969467182270 \\
$l=4$ &-1.19412096570063 & 0.75616420973758 & -0.12462741476560 & -0.01366414279716 \\
$l=5$ &-0.29957244184153 & 0.30137080814560 & -0.13593627512955 & 0.02437018661001 \\
$l=6$ &-0.03090584845561 & 0.04725296326283 & -0.02906154255314 & 0.00673101795849 \\
$l=7$ &-0.00087783536997 & 0.00342046404978 & -0.00276819278253 & 0.00074981715535 \\
$l=8$ &0.00013718255663 & 0.00010589907564 & -0.00017873779555 & 0.00006418583561 \\
\end{tabular}
\caption{Coefficients $D_{0kl}$ of the single-charm fit~(\ref{1c ansatz}) for $k$ between 4 and 7.}
\label{table:1c02}
\end{table}
\begin{table}[h]
\centering
\small
\begin{tabular}{c|c|c|c|c}
$D_{1kl}$ & $k=0$ & $k=1$ &$k=2$ & $k=3$ \\\hline
$l=0$ & 0.00037898651589 & 0.00517918021788 & 0.04112085070528 & -0.21496054486301 \\
$l=1$ & 0.00294905480007 & -0.00603971843699 & -0.21895793674020 & -1.03193199304427 \\ 
$l=2$ & -0.01351315180078 & -0.01995812054357 & -0.22812429086629 & 3.41257368948131 \\ 
$l=3$ & 0.01475940605979 & 0.00798442052509 & 0.20625870040905 & -0.22965190841787 \\ 
$l=4$ & 0.00361472380470 & -0.04512665323742 & 0.27522405293211 & -0.58912224362215 \\ 
$l=5$ & 0.00135041781442 & 0.00257101526830 & 0.03190378201141 & -0.15565929828791 \\ 
$l=6$ & 0.00018496716551 & 0.00151028010054 & -0.00216460072286 & -0.01607688792361 \\ 
$l=7$ & 0.00003136294993 & 0.00026420915586 & -0.00084445521789 & -0.00038980944792 \\ 
$l=8$ & 0.00000345408063 & 0.00001942254381 & -0.00010722136685 & 0.00009668851788 \\ 
\end{tabular}
\caption{Coefficients $D_{1kl}$ of the single-charm fit~(\ref{1c ansatz}) for $k$ between 0 and 3.}
\label{table:1c11}
\end{table}
\begin{table}[h]
\centering
\small
\begin{tabular}{c|c|c|c|c}
$D_{1kl}$ & $k=4$ & $k=5$ &$k=6$ & $k=7$ \\\hline
$l=0$ &-1.01037416347829 & 3.85777242049678 & -4.82260835024637 & 2.43237146133600 \\
$l=1$ &7.28407942435662 & -11.41492712861138 & 3.15489853106631 & 4.33118652047488 \\
$l=2$ &-6.95529309256300 & 2.97749311540542 & 2.87779195568835 & -0.44612980575120 \\
$l=3$ &-0.81288724990409 & 1.05952773710921 & 0.58597841341702 & -0.21167555956766 \\
$l=4$ &0.25138154880765 & 0.49282349653529 & -0.24333580413142 & 0.01386846240050 \\
$l=5$ &0.20204027536542 & -0.02532694776417 & -0.04866750001498 & 0.01886365093423 \\
$l=6$ &0.04112674519580 & -0.02440884452889 & -0.00083229803405 & 0.00359648817752 \\
$l=7$ &0.00367275991399 & -0.00340746780997 & 0.00055454715801 & 0.00030764562717 \\
$l=8$ &0.00017167412681 & -0.00029973464715 & 0.00010242601552 & 0.00001903450238 \\
\end{tabular}
\caption{Coefficients $D_{1kl}$ of the single-charm fit~(\ref{1c ansatz}) for $k$ between 4 and 7.}
\label{table:1c12}
\end{table}

\end{document}